\documentclass[aps,pre,floats,twocolumn,showpacs,superscriptaddress]{revtex4}

\usepackage[utf8]{inputenc}
\usepackage[svgnames]{xcolor}
\usepackage{graphicx,epsfig}
\usepackage{times}
\usepackage{amssymb,amsmath,multirow,rotate,color}
\usepackage{xcolor}

\usepackage{float}
\usepackage{hyperref}

\usepackage{tikz}

\begin{document}

\title{Virus spread versus contact tracing: two competing contagion processes}

\author{A. Reyna-Lara}
\affiliation{Department of Condensed Matter Physics, University of Zaragoza, E-50009 Zaragoza, Spain}
\affiliation{GOTHAM Lab -- BIFI, University of Zaragoza, E-50018 Zaragoza, Spain}

\author{David Soriano-Pa\~nos}
\affiliation{Department of Condensed Matter Physics, University of Zaragoza, E-50009 Zaragoza, Spain}
\affiliation{GOTHAM Lab -- BIFI, University of Zaragoza, E-50018 Zaragoza, Spain}

\author{Sergio G\'omez }
\affiliation{Departament d'Enginyeria Inform\`atica i Matem\`atiques, Universitat Rovira i Virgili, E-43007 Tarragona, Spain}

\author{Clara Granell}
\affiliation{Department of Condensed Matter Physics, University of Zaragoza, E-50009 Zaragoza, Spain}
\affiliation{GOTHAM Lab -- BIFI, University of Zaragoza, E-50018 Zaragoza, Spain}
\affiliation{Departament d'Enginyeria Inform\`atica i Matem\`atiques, Universitat Rovira i Virgili, E-43007 Tarragona, Spain}

\author{Joan T.\ Matamalas}
\affiliation{Harvard Medical School \& Brigham and Women's Hospital, Boston MA 02115, USA}

\author{Benjamin Steinegger}
\affiliation{Departament d'Enginyeria Inform\`atica i Matem\`atiques, Universitat Rovira i Virgili, E-43007 Tarragona, Spain}

\author{Alex Arenas}
\email{alexandre.arenas@urv.cat}
\affiliation{Departament d'Enginyeria Inform\`atica i Matem\`atiques, Universitat Rovira i Virgili, E-43007 Tarragona, Spain}

\author{Jes\'us G\'omez-Garde\~nes}
\email{gardenes@unizar.es}
\affiliation{Department of Condensed Matter Physics, University of Zaragoza, E-50009 Zaragoza, Spain}
\affiliation{GOTHAM Lab -- BIFI, University of Zaragoza, E-50018 Zaragoza, Spain}
\affiliation{Center for Computational Social Science (CCSS), Kobe University, Kobe 657-8501, Japan}

\date{\today}

\begin{abstract}
After the blockade that many nations suffered to stop the growth of the incidence curve of COVID-19 during the first half of 2020, they face the challenge of resuming their social and economic activity.
The rapid airborne transmissibility of SARS-CoV-2, and the absence of a vaccine, calls for active containment measures to avoid the propagation of transmission chains. The best strategy up to date, popularly known as Test-Track-Treat (TTT), consist in testing the population for diagnosis, track the contacts of those infected, and treat by quarantine all these cases.
The dynamical process that better describes the combined action of the former mechanisms is that of a contagion process that competes with the spread of the pathogen, cutting off potential contagion pathways. Here we propose a compartmental model that couples the dynamics of the infection with the contact tracing and isolation of cases. We develop an analytical expression for the effective case reproduction number $R_c(t)$ that reveals the role of contact tracing in the mitigation and suppression of the epidemics. We show that there is a trade off between the infection propagation and the isolation of cases. If the isolation is limited to symptomatic individuals only, the incidence curve can be flattened but not bended. However, if contact tracing is applied to asymptomatic individuals too, the strategy can bend the curve and suppress the epidemics. Quantitative results are dependent on the network topology. We quantify, the most important indicator of the effectiveness of contact tracing, namely its capacity to reverse the increasing tendency of the epidemic curve, causing its bending.
\end{abstract}


\maketitle
\section{INTRODUCTION}

2020 has been a year marked by the irruption of COVID-19, the worst pandemic humanity has suffered since the Spanish Flu in 1918. From the first case reported in Wuhan on December 8 \cite{who}, as of February 2021, the disease has left more than 100 million confirmed cases and more than 2.2 million deaths worldwide \cite{who2}. The lack of antiviral prophylaxis, therapeutics, or vaccines to treat or prevent COVID-19 has put social, economic, and health systems under unprecedented strain by engaging in prolonged lockdowns all over the world. Although confinement measures have been successful in bending the epidemic trajectory \cite{Chinazzi,Giordano,Arenas}, countries face the challenge of keeping the virus transmission under control while maintaining the usual socioeconomic activity \cite{Rodo}.
\smallskip

The impact of the different control policies on the spread of transmissible diseases becomes evident from the expression of the effective reproductive number, ${\cal R}(t)$, that measures the average number of contagions that an agent, infected at time $t$, makes during its infectious period \cite{rohanibook,chowell}. This number depends on diverse epidemiological, demographical and social aspects of the particular population but, in general grounds, it can be expressed as the combination of four contributions:
\begin{equation}
{\cal R}(t)=\tau\cdot k \cdot \beta\cdot \rho_S(t)\;,
\label{eq:Reff}
\end{equation}
namely; the average duration $\tau$ of the infectious period, the average number of contacts per unit time $k$, the probability of infection per contact $\beta$, and the fraction of susceptible individuals in the population at time $t$, $\rho_S(t)$. When this number takes values larger than $1$ the number of new infections will grow in time, whereas when ${\cal R}(t)<1$ the disease is on decline and, keeping this trend in time, the number of new infections will decrease until vanishing. Thus, one of the major goals when facing an epidemic outbreak is to decrease the initial reproductive number to values below the epidemic threshold ${\cal R}(t)<1$ via either pharmaceutical or non-pharmaceutical interventions.

The existence of pharmaceutical measures such as a vaccine will reach the goal by reducing the fraction of susceptibles to a number below $(\tau\cdot k \cdot \beta)^{-1}$. However, in the absence of this possibility, non-pharmaceutical interventions must come into play. This way, social-distancing \cite{LancetPH,NHB} and the use of prophylactic measures \cite{Masks1,Masks2} (such as face masks and hands hygiene) aim to reduce, the social contacts $k$ and the disease transmissibility $\beta$ , respectively, to reach ${\cal R}(t)<1$. The degree of social distancing, and consequently the reduction of our social contacts depend on the epidemic scenario. It can range from strict closures when the incidence threatens the capacity of health systems, thus requiring ${\cal R}(t)\ll 1$ values to bend the epidemic curve, to moderate restrictions (such as banning social gatherings) when incidence is small and social activity coexists with a controlled transmission (${\cal R}(t)\lesssim1$).

The application of social distancing when trying to recover the usual socioeconomic activity involves a delicate trade-off between increasing our sociality, $k$, while controlling the transmission of a virus that takes advantage of our interactions to spread. For this reason, pro-active control measures such as Test-Treat-Track (TTT) are mandatory to reach the former balance and avoid future epidemic waves \cite{ct1,ct2}. This strategy is based on the detection of symptomatic individuals, isolating them, and, more importantly, trace the contacts that these individuals have had in the recent past as they represent potential infections before the index case has been detected.  A successful contact tracing stops the spread of the virus caused by these secondary cases and hence reduces the average infectious period, $\tau$, in Eq.~(\ref{eq:Reff}) leading to a decrease of ${\cal R}(t)$.

A successful contact tracing requires a personalized and exhaustive search of the contacts of each detected case, taking into account the complex and heterogeneous nature of human relationships \cite{Makse2020TTT,Dufresne2020TTT,Rapisardi2020TTT,Barrat2020TTT}. This arduous task, however, becomes critical when, as in the case of SARS-CoV-2, pre-symptomatic and asymptomatic infections are abundant \cite{CT1,CT2,CT3,CT4,CT5,CT6,CT7}. Under these conditions, the symptomatic cases that are detected have already infected some of their contacts and, in addition, it is possible that a large fraction of their known infectees do not present symptoms during the entire infectious period.

\begin{figure}[t!]
\centering
\includegraphics[width=0.48\textwidth,angle=0]{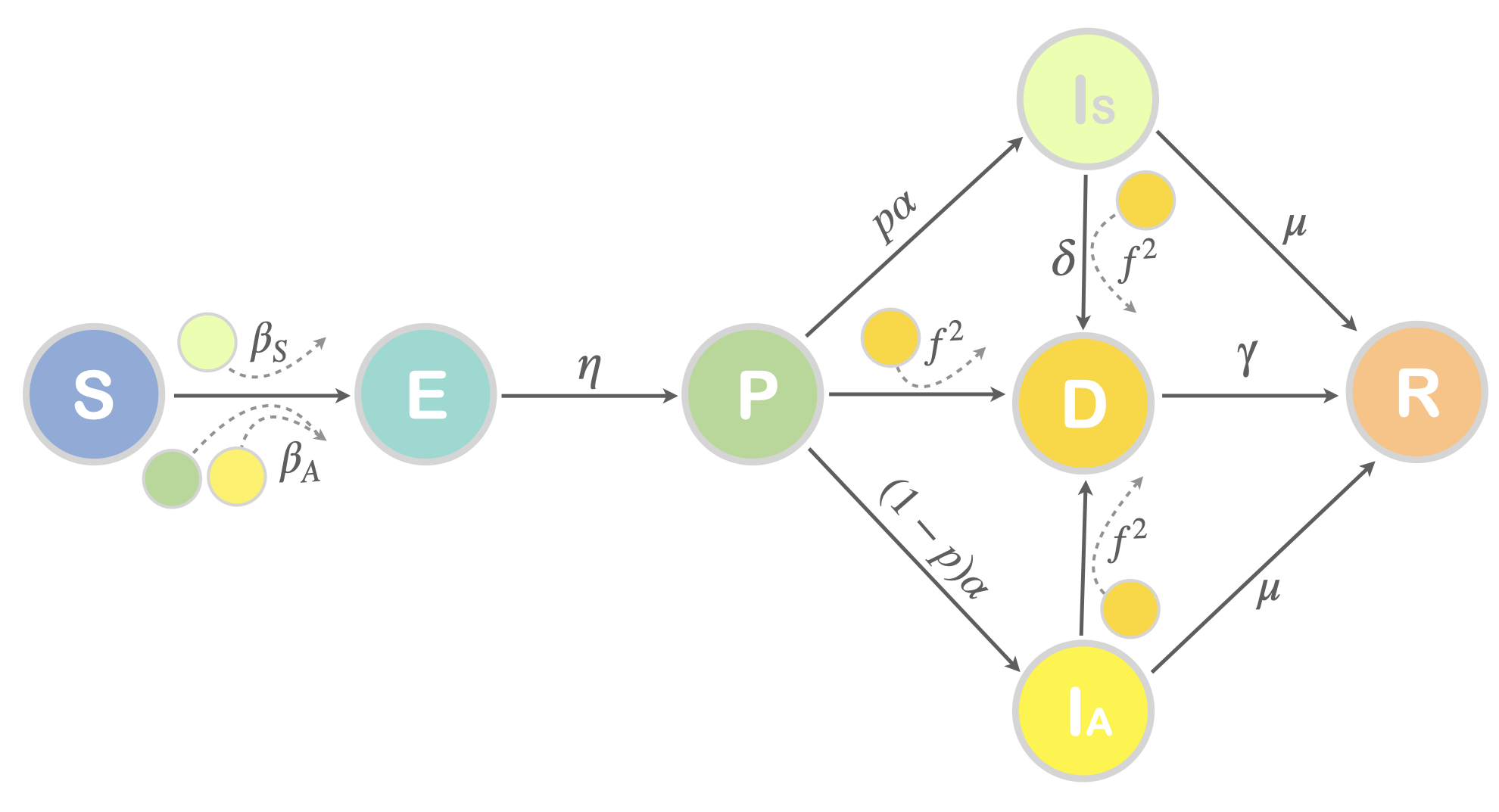}
\caption{{\bf Flow diagram of the compartmental epidemic Model.} The  epidemiological compartments are: susceptible ($S$), exposed ($E$), presyntomatic ($P$), infected asymptomatic ($I_A$), infected symptomatic ($I_S$), detected ($D$) and recovered ($R$). Arrows represent the possible transitions between the different states.}
\label{Diagram}
\end{figure}

\section{Contact-tracing as a competitive contagion process}

The essential characteristics of contact tracing (termed CT hereafter)  can be captured by modeling it as a contagion process in which the infectious agents detected spread the possible identification of other positive cases through their social network. This process competes with the spread of the pathogen itself and aims to suppress the transmission of the virus by eliminating the active spreaders whose infection is related to the identified case. Unlike other competing processes in which different viruses spread simultaneously in a population and interfere with the transmission of others \cite{Karrer,Granell,Sanz,Sahneh,Kogan,Poletto,Dufresne,Soriano}, here the cascade of CT identifications can only be activated by the presence of the pathogen.

To incorporate the CT dynamics into a framework capturing the transmission of SARS-CoV-2 we first construct an epidemic model including $7$ compartments (states): Susceptible ($S$), Exposed ($E$), Pre-symptomatic ($P$), Infectious asymptomatic ($I_A$), Infectious symptomatic ($I_S$), Detected ($D$), and Removed ($R$). The transitions between these states are shown in Fig.\ref{Diagram} and explained as follows.

Susceptible ($S$) agents are healthy individuals who can be infected by direct contact with Asymptomatic and Symptomatic Infectious agents with probability $\beta_A$ and $\beta_S$ respectively. When an $S$ agent is infected, she converts into Exposed ($E$) in which the individual displays no symptoms and is not contagious. This state lasts for an average period of $\eta^{-1}$ days. After being in $E$, agents pass to the Pre-symptomatic state ($P$). In this state, no symptoms are observed but the individual is already contagious, with contagion probability per contact of $\beta_A$. At this $P$ stage, and without detection, the individual lasts an average of $\alpha^{-1}$ days.

After the $P$ stage, individuals can continue being asymptomatic  ($I_A$) with some probability $(1-p)$ that is given by the fraction of fully asymptomatic infections. Individuals in this compartment share the same characteristics regarding infectivity and detectability as $P$. Without detection, an individual lasts an average of $\mu^{-1}$ days in this compartment before entering the Removed state ($R$). The rest (a fraction $p$) of $P$ agents pass to be Symptomatic Infectious ($I_S$). This compartment is characterized by an infectivity $\beta_S$ when contacting an $S$ agent and, as in the case of the $I_A$ compartment, and (average) of $\mu^{-1}$ days before passing to $R$, if not detected before.

With the former ingredients, the model ($SEPI_{S}I_{A}R$) can be viewed as a refined variant of the $SEIR$ class in which compartment $I$ is split in three infectious states, $P$, $I_S$ and $I_A$ to accommodate the specific contagion forms observed for the dynamics of SARS-CoV-2 \cite{Estrada,Vespignani}. This model can be used as a framework for studying the spreading of SARS-CoV-2 and to assess the impact of contention measures such as social-distancing, prophylactic behavior or strict quarantines. However, to study the influence that detection of symptomatic cases has on the transmission dynamics and, more importantly, to incorporate the possibility of tracing those infectious contacts of symptomatic individuals, a further, and fundamental, compartment capturing those infectious agents detected ($D$) is needed.

As shown in Fig.~\ref{Diagram}, compartment $D$ can be reached by agents in $P$, $I_S$ and $I_A$  states. First, those symptomatic infectious can be directly detected as they display symptoms, this happens with probability $\delta$, that is related to the average time spanned from the onset of symptoms to the availability of the test ({\it e.g.} 2 days would correspond to $\delta=0.5$). Once a symptomatic agent is detected, CT is activated. However, the possibility of tracing recent contacts is subject to the availability of information about the social activity of those detected. Here we consider that a fraction $f$ of subjects are equipped with an application that record those acquaintances that have installed it as well. Thus, those contacts of $D$ individuals that are in the $P$, $I_A$, and $I_S$ states can transit to state $D$ by means of an infection-like process in which the infection probability is equal to $f^2$, {\em i.e.} the probability that both the detected individual (in $D$) and the corresponding infectious contact (either in $P$, or $I_A$ or $I_S$) are equipped with the tracking application. Finally, any individual entering in $D$ transits to Removed ($R$) with a probability $\gamma$, {\it i.e.}, the CT contagion-like process has an {\em effective infectious period} of  $\gamma^{-1}$ days. In the following we will set $\gamma=1$ considering that, once an agent is detected, the corresponding infectious contacts are immediately identified.

The complete model has $6$ epidemiological parameters (those of the $SEPI_{S}I_{A}R$ model) and $3$ additional ones, $\delta$, $f$, and $\gamma$, that characterize the CT contagion process triggered by symptomatic detection. The values of the epidemiological parameters are presented in Table~\ref{table1} with the exception of $\beta_{\text S}$ and $\beta_{\text A}$ that are assumed to be equal, $\beta_{\text S}=\beta_{\text A}$, and whose value is taken so that the attack rate in the absence of detection, $R^{\infty}$, is the same in all the networks analyzed. Having fixed the epidemiological parameters, those corresponding to detection are used to analyze the impact of CT on epidemics.

\begin{table}[t!]
\begin{tabular}{c | c |  c | c }
 Parameter & Value & Description & Reference\\
\hline
$\eta$ & $1/2.5$ day$^{-1}$ & Probabilty $E \rightarrow P$ & \cite{timesincpre} \\
$\alpha$ & $1/2.5$ day$^{-1}$ &  Probabilty $P \rightarrow I_{A}\;, I_{S}$ & \cite{timesincpre}\\
$p$ & $0.65$ & Fraction of Symptomatic &  \cite{CDC} \\
$\mu$ & $1/7$ day$^{-1}$& Probability $I_A\;, I_S\rightarrow R$ &\cite{Masks2,infectiousperiod}\\
\end{tabular}
\caption{Epidemiological parameters of the compartmental model.}
\label{table1}
\end{table}

\subsection{Markovian dynamics}
The dynamical evolution of the compartmental model can be studied under a microscopic Markovian time-discrete formulation \cite{EPL,PRE1,PRE2}. In this framework, the dynamical state of a node $i$ at time $t$ is given by the probability of being susceptible, $\rho_i^{S}(t)$, exposed, $\rho_i^{E}(t)$, pre-symptomatic, $\rho_i^{P}(t)$, infectious asymptomatic, $\rho_i^{I_A}(t)$, infectious symptomatic, $\rho_i^{I_S}(t)$, detected, $\rho_i^{D}(t)$, and recovered, $\rho_i^{R}(t)$. The evolution of these probabilities is then given by:
\begin{widetext}
\begin{eqnarray}
\rho_i ^{\text E}(t+1)&=& (1-\eta)\rho^{\text E}_i(t)+\left(1-\rho_{i}^{\text E}(t)-\rho_{i}^{\text P}(t)-\rho_{i}^{\text I_A}(t)-\rho_{i}^{\text I_S}(t)-\rho_{i}^{\text D}(t)-\rho_{i}^{\text R}(t)\right)\Pi^{\text S \rightarrow \text E}_{i}(t) \\
\rho_i ^{\text P}(t+1)&=& (1- \Pi^{\text P \rightarrow \text D}_{i}(t)) (1-\alpha)  \rho_i ^{\text P}(t) + \eta\rho^{\text E}_i(t) \\
\rho_i ^{\text I_A}(t+1)&=& (1- \Pi^{\text I_A \rightarrow \text D}_{i}(t)) (1-\mu)  \rho_i ^{\text I_A}(t) + (1- \Pi^{\text P \rightarrow \text D}_{i}(t))(1-p)\alpha\rho^{\text P}_i(t) \\
\rho_i ^{\text I_S}(t+1)&=& (1-\Pi^{\text I_S \rightarrow \text D}_{i}(t))(1-\mu)  \rho_i ^{\text I_S}(t) +  (1- \Pi^{\text P \rightarrow \text D}_{i}(t))p\alpha \rho_i ^{\text P}(t) \\
\rho_i ^{\text D}(t+1)&=& (1-\gamma)\rho_i ^{\text D}(t) + \Pi^{\text I_S \rightarrow \text D}_{i}\rho_i ^{\text I_S}(t) + \Pi^{\text I_A \rightarrow \text D}_{i}\rho_i ^{\text I_A}(t) + \Pi^{\text P \rightarrow \text D}_{i}(t)\rho_i ^{\text P}(t) \\
\rho_i ^{\text R}(t+1)&=& \rho_i ^{\text R}(t) + \gamma\rho_i^{\text{D}}(t) + \mu(1-\Pi^{\text I_S \rightarrow \text D}_{i}(t)) \rho_i ^{\text I_S}(t)
 + \mu(1-\Pi^{\text I_A \rightarrow \text D}_{i}(t)) \rho_i ^{\text I_A}(t) \;,
\label{eq:1}
\end{eqnarray}
\end{widetext}
where we have omitted the equation for $\rho_i^{S}(t)$ due to the normalization condition: \begin{equation}
\rho_i^{S}(t)+\rho_i^{E}(t)+\rho_i^{P}(t)+\rho_i^{I_A}(t)+\rho_i^{I_S}(t)+\rho_i^{D}(t)+\rho_i^{R}(t)=1\;.
\end{equation}

In the former equations the quantities $\Pi^{\text S \rightarrow \text E}_{i}(t)$, $\Pi^{\text P \rightarrow \text D}_{i}(t)$, $\Pi^{\text I_A \rightarrow \text D}_{i}(t)$ and $\Pi^{\text I_S \rightarrow \text D}_{i}(t)$ account for the probabilities that an individual passes from Susceptible to Exposed, from Pre-symptomatic to Detected, from Infectious asymptomatic to Detected, and from Infectious symptomatic to Detected respectively. Considering the adjacency matrix ${\bf A}$ capturing the contacts between the nodes ($A_{ij}=1$ if $i$ and $j$ are connected and $A_{ij}=0$ otherwise), these probabilities read:
\begin{equation}
\Pi^{\text S \rightarrow \text E}_{i}(t)= 1- \prod_{j=1} ^N \left\{1-A_{ij}\left[\beta_{\text A}\left( \rho_{j}^{\text P}(t) + \rho_{j}^{\text I_A}(t)\right) + \beta_{\text S}\rho_{j}^{\text I_S}(t)\right]\right\},
\label{Qse}
\end{equation}
\begin{equation}
\Pi^{\text P \rightarrow \text D}_{i}(t)=\Pi^{\text I_A \rightarrow \text D}_{i}(t)= 1- \prod_{j=1} ^N \left(1-A_{ij} f^2 \rho_j^{\text{D}}(t)\right),
\label{Qad}
\end{equation}
\begin{equation}
\Pi^{\text I_S \rightarrow \text D}_{i}(t)=1-(1-\delta)\prod_{j=1} ^N \left(1-A_{ij} f^2 \rho_j^{\text{D}}(t)\right).
\label{Qsd}
\end{equation}
Note that in Eqs. (\ref{Qad})-(\ref{Qsd}) the probabilities of being detected, $\Pi^{\text P \rightarrow \text D}_{i}(t)$, $\Pi^{\text I_A \rightarrow \text D}_{i}(t)$ and $\Pi^{\text I_S \rightarrow \text D}_{i}(t)$, are calculated as $1$ minus the probability of the complement, i.e., the probability of not being detected. In the first two cases, $\Pi^{\text P \rightarrow \text D}_{i}(t)$ and $\Pi^{\text I_A \rightarrow \text D}_{i}(t)$, the probability of the complement contains the product of the probabilities of not being traced from any detected neighbor $j$, $(1-A_{ij}f^2\rho_j^{\text{D}}(t))$. In addition, the probability of the complement in $\Pi^{\text I_S  \rightarrow \text D}_{i}(t)$ also considers the probability of not being detected through symptomatic detection $(1-\delta)$.

Although the equations above only give information about the probability that a node $i$ is in the Detected compartment at each time, it is possible to construct the probability that a given node $i$ is detected at time $t$, either after showing symptoms, $D_i^{S}(t)$, or via contact tracing, $D_i^{CT}(t)$. Thus, the expected number of symptomatic detections at time $t$ is:
\begin{equation}
D^{S}(t)=\sum_{i=1}^{N}D_i^{S}(t)=\delta\sum_{i=1}^{N}\rho_i^{I_S}(t)\;,
\end{equation}
and the expected number of detections via CT at time $t$ is:
\begin{widetext}
\begin{equation}
D^{CT} (t)=\sum_{i=1}^{N}D_i^{CT}(t)=\sum_{i=1}^{N}\left\{\rho_{i}^{\text P}(t)\Pi^{\text P \rightarrow \text D}_{i}(t)+\rho_{i}^{\text A}(t)\Pi^{\text I_A \rightarrow \text D}_{i}(t)
+\rho_i^{I_S}(t)\left[(1-\delta)\left(1-\prod_{j=1} ^N \left(1-A_{ij} f^2 \rho_i^{\text{D}}(t)\right)\right)\right]\right\}\;.
\end{equation}
\end{widetext}
This last expression captures the effects of network topology on the success of CT strategies. Note that the expressions of $D^{S}(t)$ and $D^{CT}(t)$ are not used as inputs for the Markovian equations. On the contrary, they are obtained from the time evolution of the individual probabilities associated to compartments $P$, $I_S$, $I_A$ and $D$.

\begin{figure*}[t!]
\centering
\includegraphics[width=0.78\textwidth]{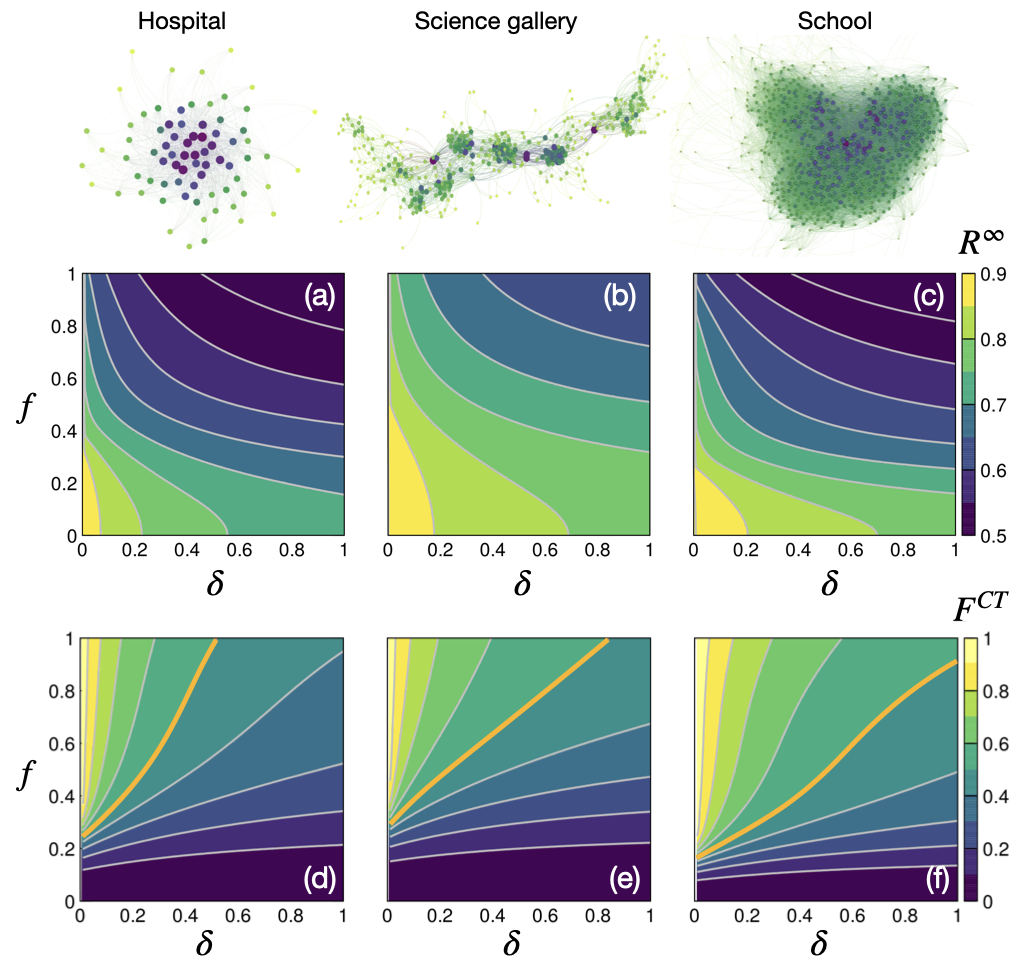}
\caption{{\bf CT versus symptomatic detection in three proximity networks (top)}. The color and size of the nodes is proportional to their connectivity, being the small (large) and yellow (blue) ones those vertices with the smallest (largest) degree. Panels (a)-(c) show the attack rate, $R^{\infty}$ as a function of the quality of symptomatic $\delta$ and CT detection $f$. Panels (d)-(f) show the fraction of detected cases by CT, $F^{CT}$. We have highlighted the case $F^{CT}=0.5$ (orange line) signalling that symptomatic and CT detections identify the same number of cases. The infectivity probability per contact ($\beta_{\text S}=\beta_{\text A}=\beta$) is chosen so that the attack rate in the absence of any kind of detection ($\delta=f=0$) is $R^{\infty}=0.9$ for the three networks.}
\label{Equilibria}
\end{figure*}

\section{Symptomatic detection versus CT}

The solution of the former Markovian equations allows to explore the performance of CT on any particular social network characterized by its adjacency matrix ${\bf A}$ in a fast and accurate way. In Appendix A, we show the validity of these equations by comparing with the results obtained through mechanistic stochastic simulations of the compartmental dynamics. In the following, we will focus on three real proximity networks with different populations and social structures in which data were obtained by means of face-to-face sensors that capture interactions with a temporal resolution of 20 s. In Table~\ref{table2} we report the main structural descriptors of these proximity networks. Although these networks can be represented as time-varying or weighted graphs, we created the static unweighted network for each case. Particularly, for the school network, we set a temporal window of 5 minutes as the minimum interaction time to define a link between individuals and focus our analysis on its giant component.

\begin{table}[b!]
\begin{tabular}{c c c c c}
Network & \footnotesize $N$ & $<k>$ & $r$ & Reference\\
\hline
Hospital & 75 & 30 & -0.18076 & \cite{Hospital}\\
Science Gallery & 410 & 13 & 0.22575 & \cite{Dublin}\\
School & 784 & 60 & 0.22814 & \cite{School} \\
\end{tabular}
\caption{{\bf Characterictics of the three proximity networks}. For each network we show the number of nodes $N$, the average degree of the nodes $\langle k\rangle$, and the assortatitivity measured as the Pearson correlation between the degrees of adjacent nodes $r$. We also report the reference were these networks where presented and analyzed.}
\label{table2}
\end{table}

Once the contact networks are constructed from proximity data we implement the compartmental model equipped with the symptomatic and CT detection to unveil the effects of these mechanisms on the spreading dynamics. In Fig.~\ref{Equilibria}.a-c we show the diagrams $R^{\infty} (\delta,f)$ for the three proximity networks analyzed. In all the plots it becomes clear that the sole implementation of symptomatic detection ($f=0$) does not lead to a dramatic decrease of the final attack rate. On the contrary, even with a poor symptomatic detection ({\it e.g.} $\delta=0.2$), the addition of CT with a moderate penetration ({\it e.g.} $f=0.5$) yields much lower attack rates than the case with perfect symptomatic detection and no CT ({\it i.e.} $\delta=1$ and $f=0$).

We analyze the combined impact of CT and symptomatic detection in panels (d)-(f) of the same Fig.~\ref{Equilibria}. There we plot the fraction of cases detected via CT with respect to the total number of infectious cases identified:
\begin{equation}
 F^{CT}(\delta, f)=\frac{\sum_{t=0}^{\infty}D^{CT}(t)}{\sum_{t=0}^{\infty}\left(D^{CT}(t)+D^{S}(t)\right)}\;.
\end{equation}
In these plots we highlight the curve (orange) corresponding to those values of $\delta$ and $f$ that yield $F^{CT}=0.5$.  Although the partition between CT and symptomatic detections depends on the precise network architecture, from the panels it becomes clear that CT alone is not responsible of the large decrease in the attack rate produced for large values of $\delta$ and $f$, but is the combination of both mechanisms what allows the efficient suppression of transmission chains.

\begin{figure}[t!]
\centering
\includegraphics[width=0.46\textwidth]{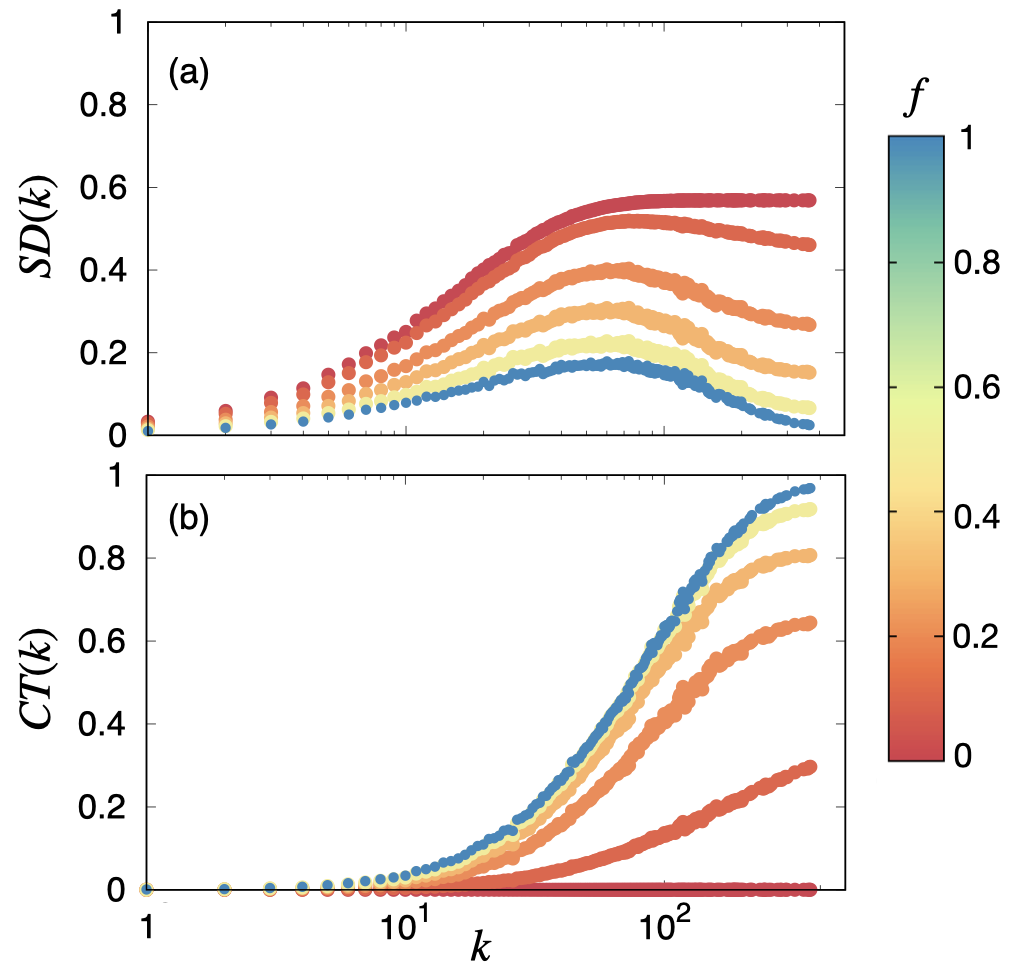}
\caption{{\bf Symptomatic detection and CT as a function of degree $k$ of the nodes}. The panel shows the probability of being detected via symptoms (a) and CT (b) as a function of the degree of the nodes for the School proximity network.The curves correspond (from red to blue) to $f=0$, $0.1$, $0.2$, $0.3$, $0.5$ and $1$.}
\label{SDk}
\end{figure}

\subsection{Microscopic differences: CT versus symptomatic detection}

The last result is quite expected since CT cannot show up alone, since it is triggered by symptomatic detections. However,
from the panels (a)-(c) in Fig.~\ref{Equilibria} it is clear that symptomatic detection alone does not allow a significant decrease of the epidemic impact but it needs the addition of CT policies. It is thus the combination of these two policies what makes detection effective. However, although the number of detections made with each of the two mechanisms is roughly similar when reaching the maximum decrease of the attack rate $R^{\infty}$, not all the identified cases are equally useful to stop the advance of the disease as we show below.

To shed light on the mechanisms behind the effectiveness of CT we analyze the connectivity pattern of those cases detected by CT and symptomatic detection. To this aim, we construct the probability that a node of degree $k$ has been detected during the course of an epidemic by symptomatic and CT detection:
\begin{eqnarray}
SD(k)&=&\frac{1}{N_k}\sum_{i;|\;k_i=k}\sum_{t=0}^{\infty}D_i^{S}(t)\;,\\
CT(k)&=&\frac{1}{N_k}\sum_{i;|\;k_i=k}\sum_{t=0}^{\infty}D_i^{CT}(t)\;,
\end{eqnarray}
where $N_k$ is the total number of nodes with degree $k$.

In Fig.~\ref{SDk} we plot the functions $SD(k)$ and $CT(k)$ when $\delta=0.5$ and $f$ varies in the range $f\in [0,1]$, {\it i.e.}, from no CT to a situation in which CT is always possible.  From panel (a) it is clear that when no CT is at work ($f=0$) the function $SD(k)$ is an increasing function of the degree, {\it i.e.} the largest the connectivity of a node the more probable that it is detected. This is clearly due to the high risk of infections of those nodes with a large connectivity that, consequently, have more probability of being at compartment $I_S$ and hence being detected. However, as $f$ increases the probability $SD(k)$ becomes a non-monotonous function of $k$ and displays a maximum at some degree class $k^{\star}$. The reason behind this behavior is the action of $CT$, that shows [see panel (b)] an increasing pattern for $CT(k)$ for any value of $f>0$. As $f$ increases the identification of those infected nodes with the largest degrees, {\it i.e.} the super-spreaders, is progressively replaced by CT in detriment of symptomatic detection. In fact, by comparing with the function $SD(k)$ for $f=0$, we notice that for $f=0.2$  CT already outperforms the ability of symptomatic detection in the identification of super-spreaders. Moreover, when $f>0.5$ the probability that a super-spreader is detected via CT is close to $1$, pinpointing that the effectiveness of CT is rooted on the identification and isolation of super-spreaders that have been in contact with those symptomatic cases previously detected.

 \begin{figure*}[t!]
\centering
\includegraphics[width=0.48\textwidth]{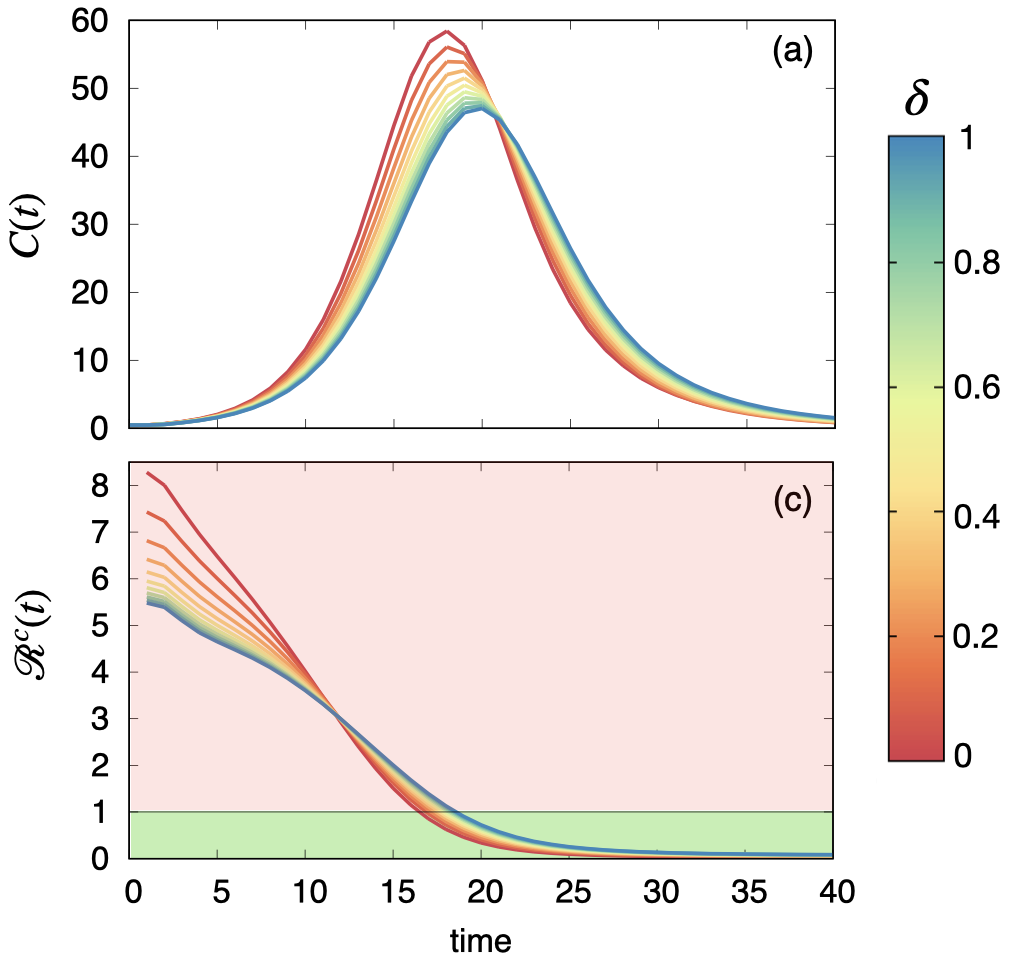}\;\;
\includegraphics[width=0.48\textwidth]{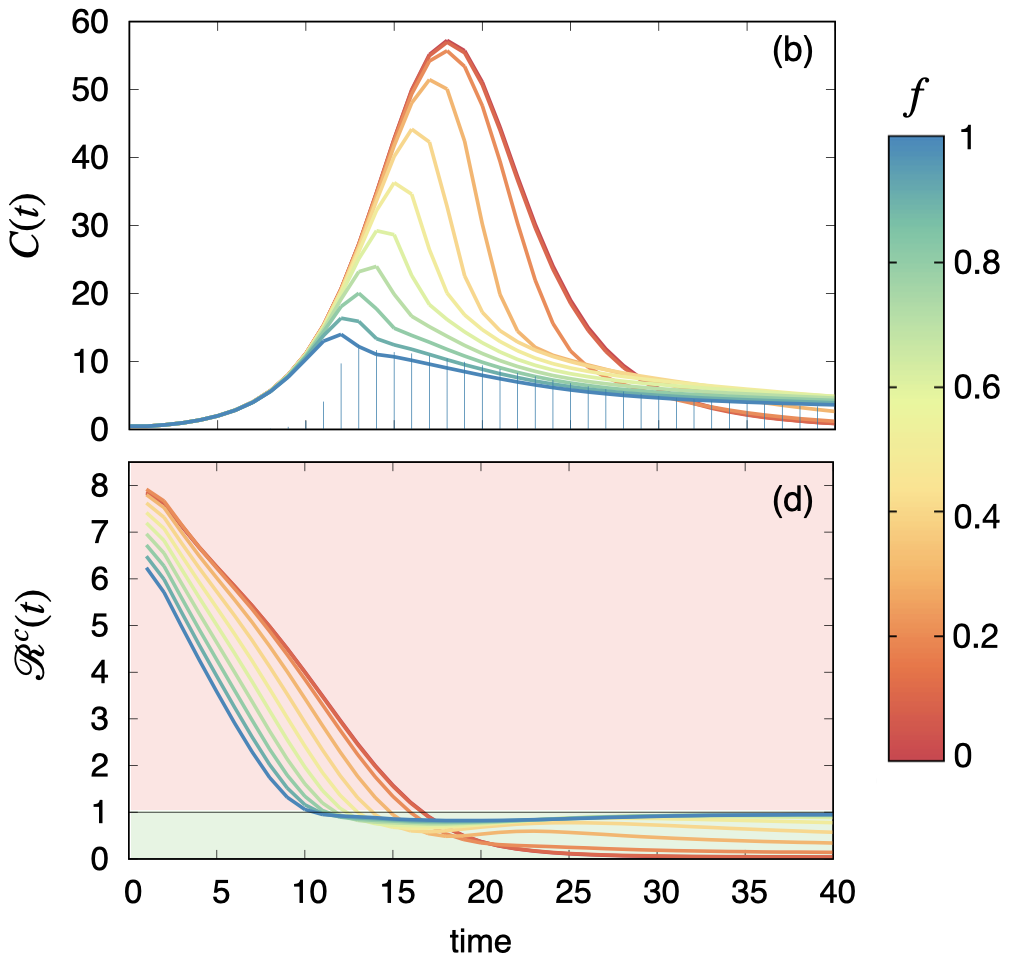}
\caption{{\bf Dynamical evolution of the epidemic trajectory under symptomatic detection and CT}.  Panels (a) and (b) show the evolution of the new contagions when different degrees of symptomatic detection and CT are implemented, respectively.  In panel (b) we also show (impulses) the evolution for number of detected cases via CT when $f=1$. The bottom panels, (c) and (d), show the evolution of ${\cal R}^{c}(t)$ corresponding to the different epidemic curves shown above. The calculations are performed using the School proximity network.}
\label{rc-te}
\end{figure*}

\subsection{Dynamical differences: Flattening vs Bending}

An early identification of super-spreaders is key to achieve an effective control of an outbreak. Super-spreaders can be identified by symptomatic detection in the first stages of an epidemic as they are usually exposed to a number of potential infections due to their large connectivity. However, symptomatic detection restricts its identification to those that display symptoms and, moreover, their identification always happens once after they have transited the $P$ compartment, thus provoking contagions in a number of neighbors prior to detection. On the contrary, CT allows catching super-spreaders at any infectious compartment, specially those in $P$, thus providing with an early suppression of super-spreading events. The earliness of CT with respect to symptomatic detection is manifested in the progressive replacement in the identification of large degree nodes as $f$ increases observed in Fig.~\ref{SDk}.

The early identification of super-spreaders provided by CT is more evident when analyzing the time evolution of the epidemic curve when subjected either to symptomatic detection or to CT. To monitor the effects that both detection mechanisms have on the epidemic trajectory we monitor in panels (a) and (b) of Fig. \ref{rc-te} the time evolution of the number of new contagions when symptomatic detection and CT are the only detection mechanisms respectively. The number of new contagions that occur at a given time $t$, $C(t)$, can be readily computed from the Markovian dynamics as:
\begin{equation}
C(t)=\sum_{i=1}^{N}\rho_i^S(t)\Pi^{S\rightarrow E}_i(t)\;.
\end{equation}

The first set of curves, panel (a), shows how symptomatic detection changes the epidemic curve as $\delta$ varies from $0$ to $1$. This plot shows that the sole action of symptomatic detection causes the so-called flattening of the epidemic curve in which the peak of the epidemic curve is delayed and decreased. This flattening becomes more pronounced as $\delta$ increases, thus reducing progressively the final attack rate. On the other hand the effect of CT, see panel (b), yields a qualitative different scenario. In this case we set a very small degree of symptomatic detection $\delta=0.05$ to trigger the CT cascade, and vary $f$ from $0$ to $1$. The result is that the epidemic curve is no longer flattened but bent, {\em i.e.} CT is able to reverse the increasing tendency of the curve corresponding to $f=0$. This bending occurs the sooner the larger the fraction $f$ of individuals adopting the CT application. To illustrate the bending action of CT we show (blue impulses) the evolution of the number of CT detections, $D^{CT}(t)$ [Eq.~(12)], for the case $f=1$. It is clear that as soon as CT is triggered the increasing trend of the epidemic curve is reversed leading to a successful mitigation.

\subsubsection{Effective case reproduction number ${\cal R}^{c}(t)$}

To shed more light on the qualitative differences between CT and symptomatic detection, we can monitor the effective case-reproductive number ${\cal R}^c(t)$ to analyze their respective impact on the evolution of the infective power of nodes in the network. ${\cal R}^{c}(t)$ is defined as the average number of secondary cases that a case infected at time step $t$ will eventually infect during her infectious period \cite{cori}. Here, an agent can transit three infectious states, namely $P$ and $I_S$ and $I_A$, thus, in general, the effective case-reproduction number would be:
 \begin{equation}
{\cal R}^{c}(t)={\cal R}^{c}_{P}(t)+p\cdot {\cal R}^{c}_{I_S}(t)+(1-p)\cdot {\cal R}^{c}_{I_A}(t)\;,
\end{equation}
where ${\cal R}^{c}_{\star}(t)$ is the average number of infections made by an agent infected at time $t$ when staying at compartment $\star$. However, when CT is active, the time window associated to each infectious compartment does depend on the instant state of the system, and the former partition is not straightforward. In this case, the calculation of ${\cal R}^{c}_{\star}(t)$ should be performed starting from its general definition:
 \begin{equation}
 {\cal R}^{c}(t)=\frac{\sum_{i=1}^{N}\rho_i^S(t-1)\Pi_i^{S\rightarrow E}(t-1){\cal{I}}_i(t)}{\sum_{i=1}^{N}\rho_i^S(t-1)\Pi_i^{S\rightarrow E}(t-1)}
 \label{RcaseGeneral}
 \end{equation}
where ${\cal{I}}_i(t)$ is the number of infections caused by agent $i$ provided she has been infected at precise time $t$.

 To calculate ${\cal{I}}_i(t)$ in a general way, we introduce the join probabilities ${\cal P}_i(\tau_E, \tau_P, \tau_{A}|t)$ and ${\cal P}_i(\tau_E, \tau_P, \tau_{S}|t)$ that account for the probabilities that an agent $i$ infected at time $t$ stays a time $\tau_E$ in the exposed compartment, a time $\tau_P$ in the pre-symptomatic stage, and a time $\tau_{A}$ or $\tau_{S}$ in the infectious asymptomatic or symptomatic stages respectively. Note that these two probabilities does not depend on $t$ and factorize,
\begin{equation}
{\cal P}_i(\tau_E, \tau_P, \tau_{{\star}}|t)={\cal P}(\tau_E){\cal P}(\tau_P){\cal P}(\tau_{{\star}})\;,
\end{equation}
when the time interval in each compartment does not depend on the state of the system, as it is the case when $f=0$. The general form of these conditional probabilities for any value of $\delta$ and $f$ probabilities is derived in Appendix B.

Once known the probabilities ${\cal P}_i(\tau_{E},\tau_{P},\tau_{S}| t)$ and ${\cal P}_i(\tau_{E},\tau_{P},\tau_{A}| t)$, the average infections made by an individual $i$ can be written as:
\begin{widetext}
\begin{eqnarray}
{\cal{I}}_i(t)&=& p\sum_{\tau_E=1}^{\infty}\sum_{\tau_P=1}^{\infty}\sum_{\tau_S=0}^{\infty}{\cal P}_i(\tau_{E},\tau_{P},\tau_{S}| t)
\left\{
\sum_{s=t+\tau_E+1}^{t+\tau_E+\tau_P}\sum_{j=1}^{N}A_{ij}\beta_{A}\rho_j^S(s)
+\sum_{s=t+\tau_E+\tau_P+1}^{t+\tau_E+\tau_P+\tau_I}\sum_{j=1}^{N} A_{ij}\beta_{I}\rho_j^S(s)
\right\}\nonumber\\
&+&(1-p)\sum_{\tau_E=1}^{\infty}\sum_{\tau_P=1}^{\infty}\sum_{\tau_A=0}^{\infty}
{\cal P}_i(\tau_{E},\tau_{P},\tau_{A}| t)\left\{
\sum_{s=t+\tau_E+1}^{t+\tau_E+\tau_P+\tau_A}\sum_{j=1}^{N}A_{ij}\beta_{A}\rho_j^S(s)
\right\}\;,
\end{eqnarray}
\end{widetext}
and the evolution of ${\cal R}^{c}(t)$ can be computed to illustrate the qualitative differences between symptomatic detection and CT.

In the bottom panels, (c) and (d), of Fig.~\ref{rc-te}  we show the evolution of ${\cal R}^{c}(t)$ for the different epidemic curves shown in panels (a) and (b). The evolution of ${\cal R}^{c}(t)$ when symptomatic detection is at work shows the fingerprint of the flattening effect observed in panel (a), {\em i.e.}, the effective reproduction number, while being smaller in the beginning of the epidemic, slows down the decreasing trend as $\delta$ increases, thus reaching  ${\cal R}^{c}(t)=1$ at larger times. On the contrary, from panel (d) we observe that the action of CT is the opposite: as $f$ increases the decreasing trend of ${\cal R}^{c}(t)$ is accelerated, thus achieving ${\cal R}^{c}(t)=1$ much sooner than in the case without detection. It is also remarkable that, in the case of large values of $f$, the long term values of the effective reproductive number remain ${\cal R}^{c}(t)\lesssim 1$. This explains the situation shown in panel (b) in which an almost-steady small number of new contagions are observed after the epidemic curve is bent, thus providing a large and slow discharge of new cases. This way, CT places the system in a kind of critical equilibrium that lasts as long as there is a large enough fraction of susceptible individuals to be infected.

\begin{figure}[t!]
\centering
\includegraphics[width=0.49\textwidth]{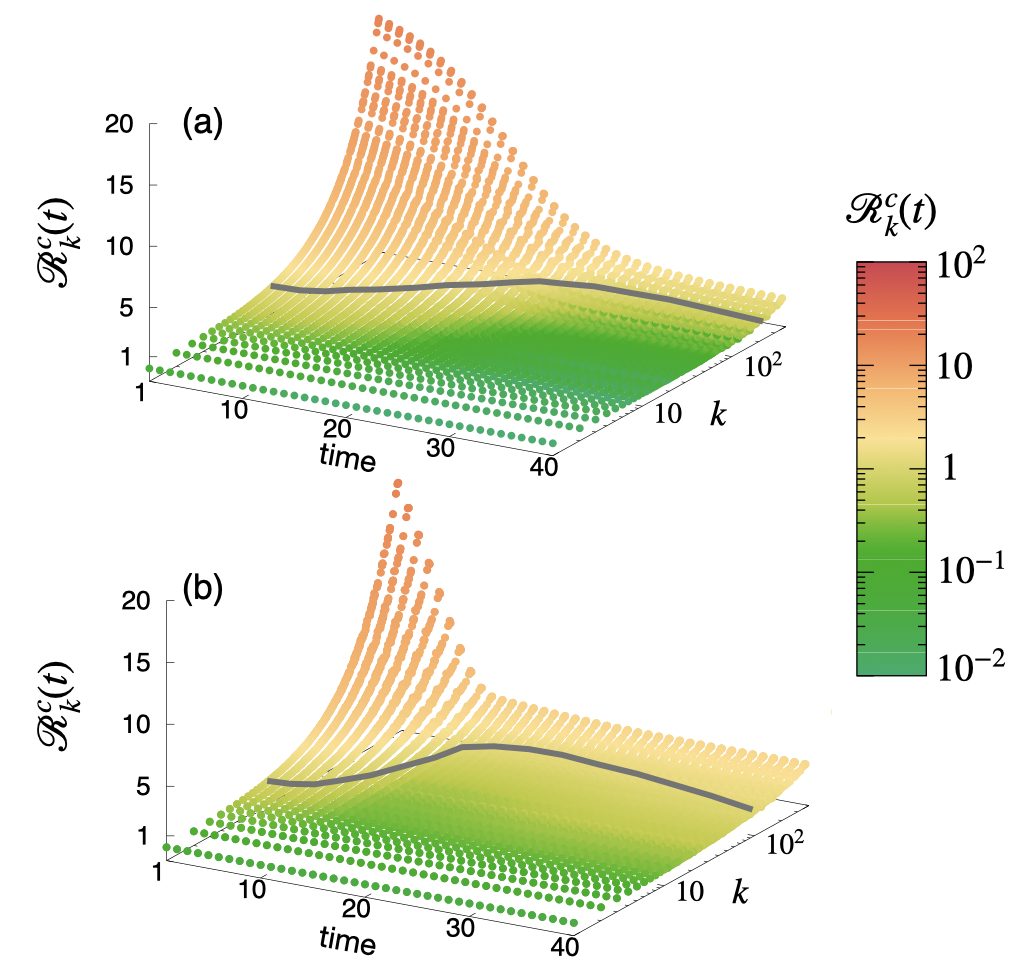}
\caption{{\bf Evolution of the case reproduction number for the different degree classes}. The panels show the evolutions of probability of  ${\cal R}^{c}_k(t)$ for (a) symptomatic detection ($\delta=1$, $f=0$) and (b) CT ($\delta=0.05$, $f=1$) for all the degree classes present in the School proximity network. The curve in grey account for the points where ${\cal R}^{c}_k(t)=1$.}
\label{5}
\end{figure}

To round off, we can use the expression for ${\cal R}^{c}(t)$ to monitor the impact on the reproduction number of each degree class. To this aim, we can define the case effective reproduction number of the nodes of degree $k$,  ${\cal R}^{c}_k(t)$, as:
 \begin{equation}
 {\cal R}^{c}_k(t)=\frac{\sum_{(i\;|k_i=k)}\rho_i^S(t-1)\Pi_i^{S\rightarrow E}(t-1){\cal{I}}_i(t)}{\sum_{(i\;|k_i=k)}\rho_i^S(t-1)\Pi_i^{S\rightarrow E}(t-1)}\;.
 \label{Rcase}
 \end{equation}
Computing this expression for each degree class, we show in Fig.~\ref{5} the evolution of ${\cal R}^{c}_k(t)$ when symptomatic detection ($\delta=1$, $f=0$) and CT ($\delta=0.05$, $f=1$) are at work in panels (a) and (b) respectively. From these two plots it becomes clear the fast drop of the infective potential of super-spreaders under the action of CT compared to the case of symptomatic detection.

\section{Conclusions}

The recent outbreaks of COVID-19 in many countries that had already controlled the first epidemic wave during the first months of 2020 show the need of efficient control measures. These measures should allow the development of normal socioeconomic activity, while avoiding the deployment of epidemic waves that threaten the sustainability of health systems. To achieve this former balance, actions aimed at eliminating local transmission chains should be implemented without jeopardizing the normal functioning of societies. To this purpose, the tracking of suspicious contacts of identified cases is key.

Here we have analyzed the effectiveness of CT by formulating its functioning as a secondary contagion dynamics that is triggered by the identification of symptomatic individuals and propagates as a detection wave. This way CT competes with the spread of the pathogen eliminating potential transmission chains. This compartmental model has been analyzed under a Markovian framework that nicely agrees with mechanistic simulations and allows a systematic study of particular network architectures such as the proximity networks and the analysis of microscopic dynamical patterns. Under this approach, we have been able to derive the evolution of the effective case reproductive number, an indicator that quantifies the average number of secondary cases that a case will eventually infect during her infectious period and, therefore, it allows to monitor both the spread of diseases and the quality of the contention strategies implemented.

Our results identify the importance of implementing CT in addition to symptomatic detection. The most important indicator of the effectiveness of CT is its capacity to sharply reverse the increasing tendency of the original epidemic curve, causing its bending. In contrast, the implementation of symptomatic detection, yields a softer modification to the epidemic trajectory, known as flattening, in which the epidemic peak is delayed and lowered.

The qualitative differences in the performance of CT and symptomatic detection are rooted microscopically. We have shown that CT allows an early detection of large degree infectious nodes, well before than symptomatic detection does. This early identification of super-spreaders during their contagious cycle is fundamental to pre-empting the spread of the virus, cut the newest transmission channels, and cause the bending of the epidemic trajectory rather than its flattening. In more general grounds, the advantage of CT to advance the, often covert, transmission of SARS-CoV-2 lies in the use of its same strategy: a fast propagation across our social fabric taking advantage of its heterogeneous nature.

Apart from these findings the formalism presented here allows to understand and quantify the impact of CT strategies in particular proximity networks that are critical to protect. Examples of these networks include companies, hospitals and schools to name a few. It also allows the possibility of designing modifications of these social structures in order to both decrease the impact of potential virus transmission and enhance the efficiency of CT strategies.

The simple nature of the Markovian model has allowed to analyze the importance of CT and the qualitative changes with respect to symptomatic detection. However, the compartmental dynamics and the social structure can be expanded in order to address the performance of CT on realistic scenarios. For instance, here we assume that CT is implemented by means of an application whose penetration is characterized by $f$. Instead, direct contacts such as those within the household or close acquaintances at workplaces can be directly tracked without the need of digital tracing. This duality can be captured with the use of multiplex social structures in which usual contacts are distinguished from casual ones. Also, here we have not considered the waiting times associated to detection. However, in many practical situations tests cannot be done immediately and thus the waiting time cooperates with the infectious period of infectious acquaintances, since they are not identified until the symptomatic case is confirmed. In addition, once confirmed, all the contacts of the detected agent are, in principle, suspects of being infectious. In those situations all the individuals are quarantined so that also Susceptible and Exposed neighbors are also removed from the population together with those Presymptomatic, Asymptomatic and Symptomatic. Another limitation is the absence of directionality of the model since here traced contacts can correspond to either secondary (forward CT) or prior (backward CT) infections of those detected cases, provided the traced cases are infective. These and other features should be considered as extensions of the model presented here.

\begin{figure*}[t!]
\centering
\includegraphics[width=0.9\textwidth]{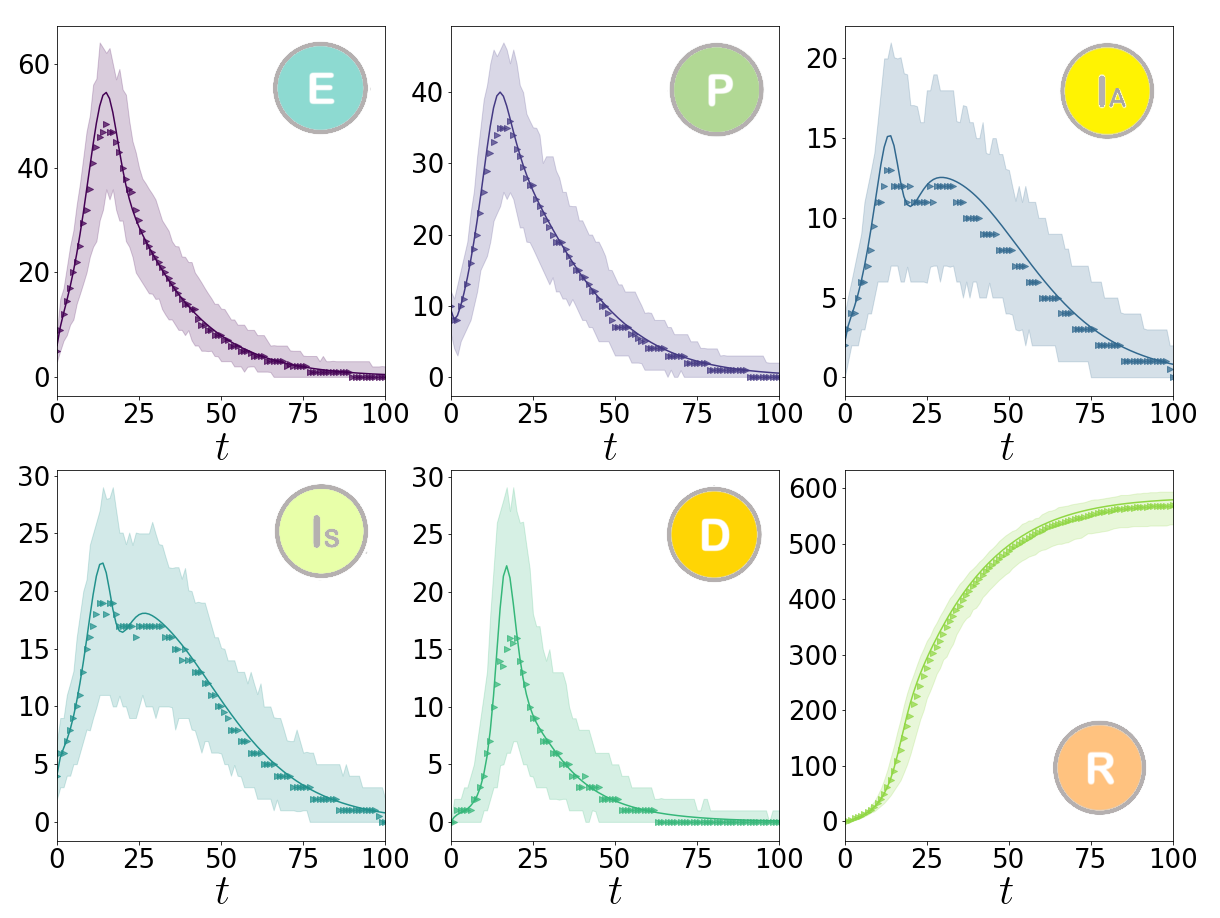}
\caption{{\bf Validation of Markovian dynamics by stochastic simulations.} Each panel shows the evolution of the occupation of each compartment as obtained from stochastic simulations (bands and points (median)) and the Markovian equations (lines).  The network substrate is the School proximity network and the detection parameters are set to $\delta=0.1$ for symptomatic detection, and $f=0.4$ for CT.}
\label{validation}
\end{figure*}

\acknowledgments
A.R.L. acknowledges support from CONACYT. A.R.L., D.S.P. and J.G.G.\ acknowledge financial support from MINECO and FEDER funds (grant FIS2017-87519-P) and from the Departamento de Industria e Innovaci\'on del Gobierno de Arag\'on y Fondo Social Europeo (FENOL group E36-20R). A.A., B.S.\ and S.G.\ acknowledge financial support from Spanish MINECO (grant PGC2018-094754-B-C21), Generalitat de Catalunya (grant No.\ 2017SGR-896), and Universitat Rovira i Virgili (grant No.\ 2019PFR-URV-B2-41). A.A.\ also acknowledges support from Generalitat de Catalunya ICREA Academia, and the James S. McDonnell Foundation (grant \#220020325).  C.G.\ acknowledges financial support from Juan de la Cierva-Formaci\'on (Ministerio de Ciencia, Innovaci\'on y Universidades). B.S.\ acknowledges financial support from the European Union's Horizon 2020 research and innovation program under the Marie Sk\l{}odowska-Curie grant agreement No.\ 713679 and from the Universitat Rovira i Virgili (URV).

\section*{APPENDIX A}

To confirm the validity of the Markovian equations we have compared the results obtained from these equations with mechanistic stochastic simulations. In these simulations we start by assigning the CT application randomly to a fraction $f$ of the agents. As initial condition we set a small fraction ($1\%$) of individuals in the $E$ compartment while the rest of the population is $S$. Then the stochastic dynamics is iterated following, at each time step, the transition rules described in Fig.~1. For each value of $f$ and $\delta$ we make $10^3$ realizations to calculate the corresponding averages.

In Fig.~\ref{validation} we compare the time evolution of the occupation of each compartment given by the iteration of the Markovian equations with the  average evolution given by the different realizations of the stochastic dynamics of the compartmental model for $\delta=0.1$ and $f=0.4$. We also plot, for each evolution, the $95\%$ CI obtained from the pool of stochastic simulations. In all the cases the Markovian equation reproduces well the trajectories obtained from mechanistic simulations, showing the accuracy of the Markovian framework used along the manuscript.

\section*{APPENDIX B}

Here we complete the derivation of the effective reproduction number ${\cal R}^c(t)$ from the Markovian equations. To this aim we show the calculation of the probability that an individual $i$ infected at time $t$ spend times $\tau_E$, $\tau_P$, and $\tau_S$ or $\tau_A$ in compartments $E$, $P$, and $I_A$ or $I_A$ respectively. Considering the Markovian equations the probability for those transiting the symptomatic phase is:
\begin{widetext}
\begin{eqnarray}
{\cal P}_i(\tau_{E},\tau_{P},\tau_{S} | t)&=&(1-\eta)^{\tau_E-1}\eta \nonumber\\
&\times&
\left[\delta_{\tau_{S},0}\cdot \Pi_{i}^{CT}(t+\tau_E+\tau_P)+(1-\delta_{\tau_S,0})\cdot\alpha\left(1-\Pi_{i}^{CT}(t+\tau_E+\tau_P)\right)\right]
\prod_{s=t+\tau_E+1}^{t+\tau_E+\tau_P-1}(1-\alpha)\left[1-\Pi_{i}^{CT}(s)\right]\nonumber
\\
&\times&
\Biggl\{\delta_{\tau_{S},0}+(1-\delta_{\tau_{S},0})\nonumber\\
&\times&
\left\{\delta + (1-\delta)\left[\Pi_{i}^{CT}(t+\tau_{T})+\left(1-\Pi_{i}^{CT}(t+\tau_{T})\right)\mu\right]\right\}\prod_{s=t+\tau_E+\tau_P+1}^{t+\tau_E+\tau_P+\tau_{S}-1}(1-\mu)(1-\delta)\left[1-\Pi_{i}^{CT}(s)\right]
\Biggr\}\;,
\label{eqPI}
\end{eqnarray}
while for those asymptomatic reads:
\begin{eqnarray}
{\cal P}_i(\tau_{E},\tau_{P},\tau_{A} | t)&=&(1-\eta)^{\tau_E-1}\eta \nonumber \\
&\times&
\left[\delta_{\tau_{A},0}\cdot \Pi_{i}^{CT}(t+\tau_E+\tau_P)+(1-\delta_{\tau_{A},0})\cdot\alpha\left(1-\Pi_{i}^{CT}(t+\tau_E+\tau_P)\right)\right]
\left[\prod_{s=t+\tau_E+1}^{t+\tau_E+\tau_P-1}(1-\alpha)\left[1-\Pi_{i}^{CT}(s)\right]\right]\nonumber \\
&\times&
\left\{\delta_{\tau_{A},0}+(1-\delta_{\tau_{A},0})\left[\Pi_{i}^{CT}(t+\tau_{T})+\left(1-\Pi_{i}^{CT}(t+\tau_{T})\right)\mu\right]
\prod_{s=t+\tau_E+\tau_P+1}^{t+\tau_E+\tau_P+\tau_{A}-1}(1-\mu)\left[1-\Pi_{i}^{CT}(s)\right]
\right\}\;,
\label{eqPA}
\end{eqnarray}
 \end{widetext}
where $\tau_T$ is defined as the total duration of the infectious period of the infected agent, {\em i.e.} $\tau_T= \tau_E +\tau_P + \tau_S$ for symptomatic individuals and $\tau_T= \tau_E +\tau_P + \tau_A$ for asymptomatic patients. In both expressions $\Pi_{i}^{CT}(t)$ is the probability of being detected through CT and is equal to $\Pi_{i}^{P\rightarrow D}(t)$ and $\Pi_{i}^{I_A \rightarrow D}(t)$ as written in Eq.~(10).

Computing Eqs.~(\ref{eqPI}) and (\ref{eqPA}) requires to save the time evolution along the epidemic trajectory of the following quantities: $\Pi_{i}^{CT}(t)$ and $\rho_i^S(t)$ for all the nodes ($i=1$,...,$N$). These two sets of quantities, together with the evolution of the infection probabilities of each node $\Pi_{i}^{S\rightarrow E}(t)$, allow us to obtain the effective case-reproduction number, ${\cal R}^{c}(t)$ in Eq.~(\ref{Rcase}).

To illustrate the validity of the former expressions, lets suppose that CT is absent, so that $\Pi_{i}^{CT}(t)=0$ $\forall$ $t$ and $i$. Then the conditional probabilities are identical for all the nodes:
\begin{eqnarray}
{\cal P}(\tau_{E},\tau_{P},\tau_{S} | t)&=&\left[\delta + (1-\delta)\mu\right](1-\mu)^{\tau_{S}-1}(1-\delta)^{\tau_{S}-1}
\nonumber\\
&\times&
\alpha(1-\alpha)^{\tau_P-1}
(1-\eta)^{\tau_E-1}\eta\;,
\end{eqnarray}
and
\begin{eqnarray}
{\cal P}(\tau_{E},\tau_{P},\tau_{A} | t)&=&\mu(1-\mu)^{\tau_{A}-1}
\alpha(1-\alpha)^{\tau_P-1}\nonumber\\&\times&
(1-\eta)^{\tau_E-1}\eta\;,
\end{eqnarray}
and the factorization of the two probabilities and their time independence becomes clear.

\end{document}